\documentclass[aps,prb,twocolumn,superscriptaddress,showpacs,showkeys,10pt]{revtex4-1}
\usepackage{bm}
\usepackage{graphicx}
\usepackage{amsmath}
\usepackage{epstopdf}

\begin{document}


\title{Crystal field investigation in the light rare earth R$_3$Pt$_{23}$Si$_{11}$ compounds.}\label{crystalelectricfield}


\author{R.M. Gal\'era}
\email{rose-marie.galera@neel.cnrs.fr}
\affiliation{CNRS, Institut N\'EEL, BP 166, F-38042 Grenoble Cedex 9, France}
\author{C. Opagiste}
\author{M. Amara}
\affiliation{Univ. Grenoble Alpes, Institut N\'EEL, BP 166, F-38042 Grenoble Cedex 9, France}
\author{J.C. Toussaint}
\affiliation{Univ. Grenoble Alpes, Institut N\'EEL, BP 166, F-38042 Grenoble Cedex 9, France}
\author{M. B. Lepetit }
\affiliation{CNRS, Institut N\'EEL, BP 166, F-38042 Grenoble Cedex 9, France}
\author{S. Rols}
\affiliation{Institut Laue-Langevin, 6 rue Jules Horowitz, BP 156, 38042 Grenoble Cedex 9, France }


\date{\today}

\begin{abstract} 
The crystalline electric field (CEF) is investigated in Pr$_3$Pt$_{23}$Si$_{11}$ and Nd$_3$Pt$_{23}$Si$_{11}$ by neutron spectroscopy (NS). At low temperature, the number of observed CEF excitations is consistent with the orthorhombic symmetry at the rare earth site. This agrees with previous results on Ce$_3$Pt$_{23}$Si$_{11}$. For Pr- and Nd$_3$Pt$_{23}$Si$_{11}$, the number of CEF parameters is too large to allow for an unambiguous determination. This determination is possible for Ce$_3$Pt$_{23}$Si$_{11}$, due to a reduced number of parameters and to the availability of extensive experimental data. A specific procedure is developed for this purpose that combines genetic algorithmics and optimization methods. An unique set of CEF parameters is found for Ce$_3$Pt$_{23}$Si$_{11}$. It reveals a strong anisotropy at the orthorhombic site, responsible for an easy threefold magnetization axis in the cubic system. Using a microscopic, mean-field, description, the magnetization processes in the paramagnetic and ferromagnetic phases of Ce$_3$Pt$_{23}$Si$_{11}$ are well reproduced. Ce$_3$Pt$_{23}$Si$_{11}$ is shown to realize a model for systems where conflicting anisotropies are forced to cooperate.

\end{abstract}


\maketitle

\section{Introduction}

We succeeded recently in synthesizing the entire series of new ternary R$_3$Pt$_{23}$Si$_{11}$ compounds where R is a lanthanide element.\cite{opagiste2012,opagiste2014,opagiste2015} They crystallize within the same face centered cubic structure ($Fm\bar3m$ space group) as reported for the first time by Tursina et \textit{al.} for Ce$_3$Pt$_{23}$Si$_{11}$.\cite{tursina2002} The evolution of the lattice parameter in the series is consistent with the lanthanide contraction effect, except for Eu$_3$Pt$_{23}$Si$_{11}$ and Yb$_3$Pt$_{23}$Si$_{11}$ where the rare earth ion is divalent. As a consequence the Eu ions bear a magnetic moment, while those of Yb are not magnetic. Joint studies of the magnetic and thermodynamic properties show that almost all the R$_3$Pt$_{23}$Si$_{11}$ present a magnetic ordering at low temperature. The ordered phase is ferromagnetic for Ce-, Sm-, Eu-, Gd-, Tb-, Dy-, Ho- and Er$_3$Pt$_{23}$Si$_{11}$ and antiferromagnetic for Nd$_3$Pt$_{23}$Si$_{11}$.\cite{opagiste2013} The ordering temperatures of compounds with trivalent rare earths follow the de Gennes law, as expected when the magnetic interactions are mediated via the indirect RKKY-type interactions. The reduced magnetic moment observed in the ordered phases for the $L\neq0$ ions, the low-temperature Van Vleck-type paramagnetism in the Pr and Tm compounds, evidence important crystalline electric field (CEF) effects.\\
Extensive studies of the magnetic properties of Ce$_3$Pt$_{23}$Si$_{11}$ were performed on a high-quality single crystal sample. The magnetic susceptibility is isotropic in the paramagnetic phase, as expected for a cubic compound, while in the ferromagnetic phase (T$_C$= 0.44 K) one observes an easy magnetization axis along the [111] direction of the cube.\cite{opagiste2011} Neutron diffraction experiments, performed on the same single crystal, reveal a quite singular spin arrangement of the six Ce ions in the unit cell. They divide into pairs, each pair having its moments aligned with one of the three fourfold axes of the cubic structure. The combination of these three directions leads to a magnetization along a threefold direction. The CEF investigation in Ce$_3$Pt$_{23}$Si$_{11}$ by neutron spectroscopy (NS), reveals two magnetic excitations at 139$\pm$5 K (12$\pm$0.5 meV) and 227$\pm$2 K (19.6$\pm$0.2 meV).\\
New NS studies have been carried out on the Pr$_3$Pt$_{23}$Si$_{11}$ and Nd$_3$Pt$_{23}$Si$_{11}$ compounds. We report here the experimental results and compare them with those of Ce$_3$Pt$_{23}$Si$_{11}$. The first section describes the experimental details. In the second section the NS results are presented. The last section is devoted to the analysis of the experimental results and to the conclusion.\\

\section{Experimental}

High-quality polycrystalline samples of Pr$_3$Pt$_{23}$Si$_{11}$ and Nd$_3$Pt$_{23}$Si$_{11}$ were prepared for neutron experiments. The stoichiometric proportions of the different constituents: Pr or Nd (99.99\%, Johnson Matthey), Pt (99.95\%, Alfa Aesar) and Si (99.9999\%, Alfa Aesar), were melted by induction technique in a cold copper crucible under a high purity argon atmosphere. Samples were melted several times to improve the homogeneity. Mass losses during this first step were less than 0.1\%. The sample quality was checked by the conventional X-ray powder diffraction technique using the Cu-K$\alpha$ radiation on a Philipps PW1730 diffractometer. Diffraction patterns are consistent with the face-centered cubic structure ($Fm\bar{3}m$ space group) and confirm that, within the experimental accuracy, no impurity phases are present.\\

Inelastic neutron scattering experiments were carried out at the Institute Laue-Langevin (ILL) in Grenoble on the IN4C time-of-flight spectrometer. Measurements were performed on two samples of 2.772 g and 2.785 g of Pr$_3$Pt$_{23}$Si$_{11}$ and Nd$_3$Pt$_{23}$Si$_{11}$ respectively. The sample holder used for these experiments consisted of a thin aluminum foil, thus reducing the contribution of the empty cell to a minimum. Three incident wavelengths, $\lambda_i$, have been selected in order to investigate the excitations over an extended energy range. The associated instrumental resolutions are determined by the full width at half maximum (FWHM) of the incoherent elastic peak: $\lambda_i$ = 1.493 {\AA} and incident energy E$_i$ = 36.7 meV with FWHM = 1.54 meV, $\lambda_i$ = 2.22 {\AA} and  E$_i$ = 16.6 meV with FWHM = 0.83 meV, $\lambda_i$ = 2.98 {\AA} and E$_i$ = 9.21 meV with FWHM = 0.35 meV. The spectra were collected in the temperature range 4 K to 150 K for scattering angles ranging from 13$^{\circ}$ to 135$^{\circ}$, and were normalized, respectively, to the incident flux and to a vanadium standard. In order to highlight the dependence on the scattering vector \textit{Q}, spectra were further averaged out in three groups, the mean scattering angles $\theta$ of which are 31.81$^{\circ}$, 67.16$^{\circ}$ and 102.77$^{\circ}$, respectively. The scattering by magnetic excitations is enhanced at low \textit{Q} values whereas phonon excitations are dominant at high \textit{Q}. Also the scattering by magnetic excitations is stronger at low temperature, while phonon scattering becomes preponderant at high temperature.\\

\section{Neutron spectroscopy}

\begin{figure}
\includegraphics[width=\columnwidth]{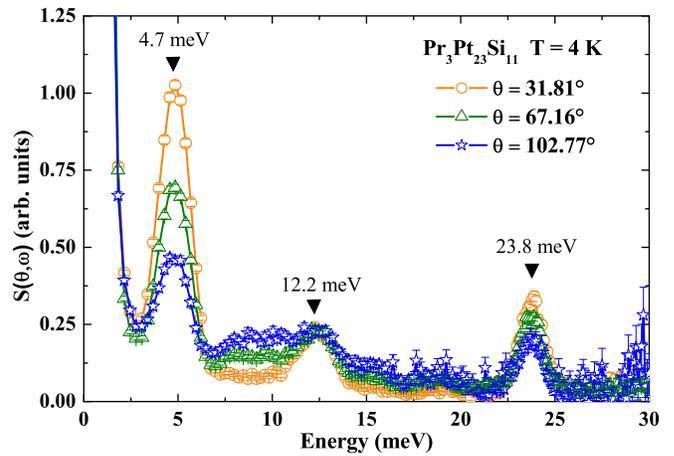}
\caption{\label{fig1} (Color online) Pr$_3$Pt$_{23}$Si$_{11}$, incident neutron energy E$_i$ = 36.7 meV: evolution of the down-scattering processes as a function of the scattering angle at T = 4 K. Dots represent the spectrum at $\theta$ = 31.81$^{\circ}$, triangles the spectrum at $\theta$ = 67.16$^{\circ}$ and stars the spectrum at $\theta$ = 102.77$^{\circ}$. Black arrows indicate the positions of the magnetic excitations at E$_1$ = 4.7 meV, E$_2$ = 12.2 meV and E$_3$ = 23.8 meV.}
\end{figure}

\begin{figure}
\includegraphics[width=\columnwidth]{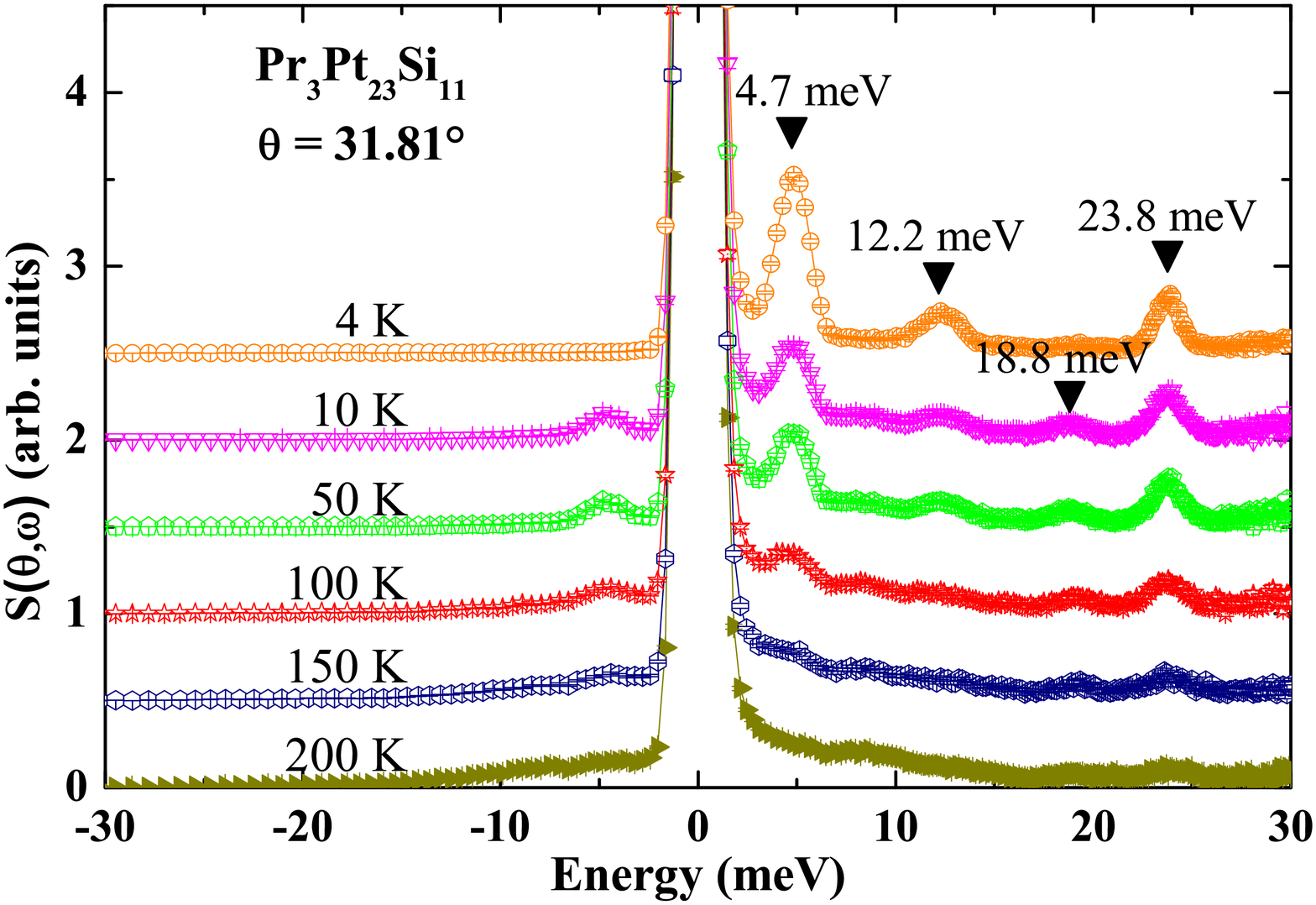}
\caption{\label{fig2} (Color online) Pr$_3$Pt$_{23}$Si$_{11}$, incident neutron energy E$_i$ = 36.7 meV: thermal evolution of the inelastic spectra. For clarity the spectra are vertically shifted by 0.5. Arrows show the positions of the magnetic excitations.}
\end{figure}

\begin{figure}
\includegraphics[width=\columnwidth]{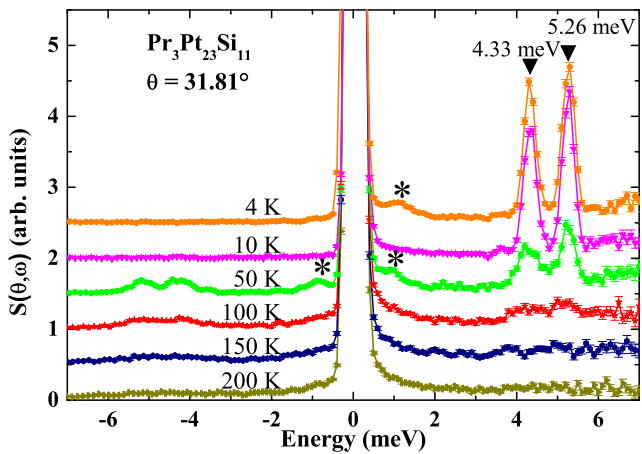}
\caption{\label{fig3} (Color online) Pr$_3$Pt$_{23}$Si$_{11}$, incident neutron energy E$_i$ = 9.21 meV: thermal evolution of the inelastic spectra. The weak structures pointed out by stars have an incoherent thermal evolution and are ascribed to spurious effects. For clarity the spectra are vertically shifted by 0.5.}
\end{figure}

For the Pr$_3$Pt$_{23}$Si$_{11}$ compound, figure~\ref{fig1} compares the three angular groups of inelastic spectra collected at T = 4 K with E$_i$ = 36.7 meV. The spectrum at $\theta$ = 31.81$^{\circ}$ shows three excitations centered at 4.7$\pm$0.2 meV (FWHM = 2.04 meV), 12.2$\pm$0.5 meV (FWHM = 2.50 meV) and E$_3$ = 23.8$\pm$0.2 meV (FWHM = 1.5 meV). Note that the FWHM of the two first excitations are larger than the experimental resolution. This may indicate that several excitations, too close in energy to be resolved, contribute to the peak. The progressive decrease of the intensity of the peaks at 4.7 meV and 23.8 meV in the spectra at $\theta$ = 67.16$^{\circ}$ and $\theta$ = 102.77$^{\circ}$ is consistent with the evolution expected for magnetic scattering. As shown in figure~\ref{fig1} the excitation at 12.2 meV is located in an energy region highly-populated with phonons. Therefore the intensity decrease at high \textit{Q} is partially masked by the increase of the phonon intensity. Very likely the peak may contain both, phonon and magnetic scattering. The magnetic origin of the peak at 12.2 meV is however indisputably confirmed in figure~\ref{fig2}, which illustrates the thermal evolution of the inelastic spectra at $\theta$ = 31.81$^{\circ}$. At this angle the phonon contribution is negligible and one observes the progressive decrease of the intensity for the three peaks with increasing the temperature. In the figure the arrows indicate the positions of magnetic excitations. At 10 K a fourth excitation appears at 18.8$\pm$0.2 meV (FWHM = 1.9 meV). Warming up the sample, its intensity progressively decreases in a similar way to the other magnetic excitations. We have pointed out the large width of some magnetic excitations therefore higher resolution spectra have been collected with E$_i$ = 9.21 meV. They are reported in figure~\ref{fig3} where two magnetic excitations can be distinguished at 4.33$\pm$0.05 meV (FWHM = 0.4 meV) and 5.26$\pm$0.05 meV (FWHM = 0.4 meV). Consequently the excitation at 18.8 meV in figure~\ref{fig2} corresponds to the transition from the second excited level at 5.26 meV to the level at 23.8 meV.

\begin{figure}
\includegraphics[width=\columnwidth]{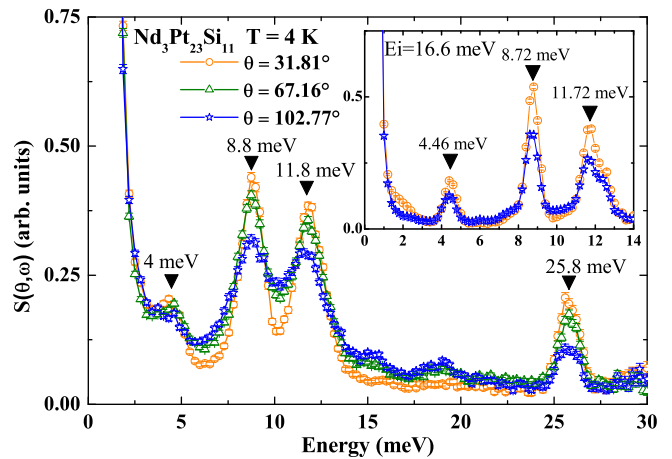}
\caption{\label{fig4} (Color online) Nd$_3$Pt$_{23}$Si$_{11}$, incident neutron energy E$_i$ = 36.7 meV: evolution of the down-scattering processes as function of the scattering angle at T = 4 K. Dots represent the spectrum at $\theta$ = 31.81$^{\circ}$, triangles the spectrum at $\theta$ = 67.16$^{\circ}$ and stars the spectrum at $\theta$ = 102.77$^{\circ}$. Black arrows indicate the positions of the magnetic excitations at E$_1$ = 4.46 meV, E$_2$ = 8.75 meV and E$_3$ = 11.72 meV and E$_4$ = 25.80 meV. The inset shows the spectra obtained with a better resolution (E$_i$ = 16.6 meV). They definitely confirm the magnetic excitation at E$_1$ = 4.46 meV.}
\end{figure}

\begin{figure}
\includegraphics[width=\columnwidth]{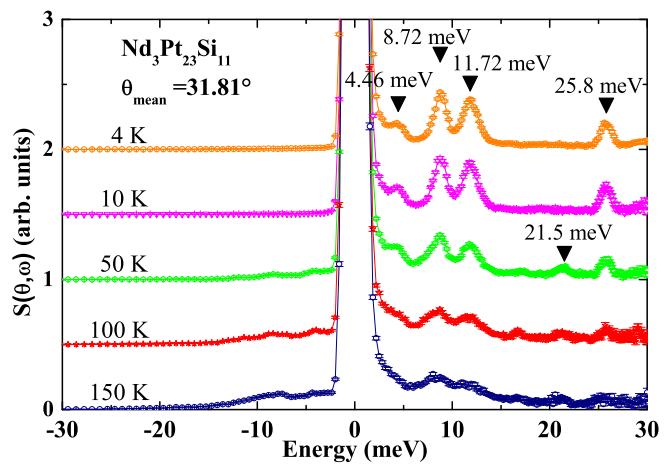}
\caption{\label{fig5} (Color online) Nd$_3$Pt$_{23}$Si$_{11}$, incident neutron energy E$_i$ = 36.7 meV: thermal evolution of the inelastic spectra. For clarity the spectra are vertically shifted by 0.5. Arrows show the positions of the magnetic excitations.}
\end{figure}

For the Nd$_3$Pt$_{23}$Si$_{11}$ compound, the evolution of the inelastic spectra as function of the scattering angle and as function of the temperature are displayed in figures~\ref{fig4} and~\ref{fig5} respectively. Spectra in figure~\ref{fig4} reveal, at 4 K, three magnetic excitations at 8.8$\pm$0.2 meV (FWHM = 1.6 meV), 11.8$\pm$0.5 meV (FWHM = 1.9 meV) and 25.8$\pm$0.2 meV (FWHM = 1.4 meV) and a bump around 4 meV. In order to resolve this bump, spectra have been collected at 4 K with $E_i$ = 16.6 meV. As shown in the inset in figure~\ref{fig4}, a well defined peak is located at 4.46$\pm$0.10 meV (FWHM = 0.73 meV). Its evolution with the scattering angle confirms a magnetic origin. In the same figure, the profile and the large FWHM of the peak centered at 11.72$\pm$0.15 meV let suppose that the peak is double. It was seen in Pr$_3$Pt$_{23}$Si$_{11}$ that phonon scattering is important in this energy range (8 - 16 meV). As the phonon scattering in Nd$_3$Pt$_{23}$Si$_{11}$ should not be very different from that in Pr$_3$Pt$_{23}$Si$_{11}$, a weak phonon contribution is very likely. In figure~\ref{fig5} the thermal evolution of the intensity of these four peaks appears fully consistent with that expected for magnetic scattering. At 50 K a fifth peak emerges at 21.5$\pm$0.2 meV (FWHM = 1.5 meV) and progressively decreases with temperature. It can be ascribed to the transition between the first excited level at 4.46 meV and the level at 25.8 meV.\\

\begin{figure}
\includegraphics[width=\columnwidth]{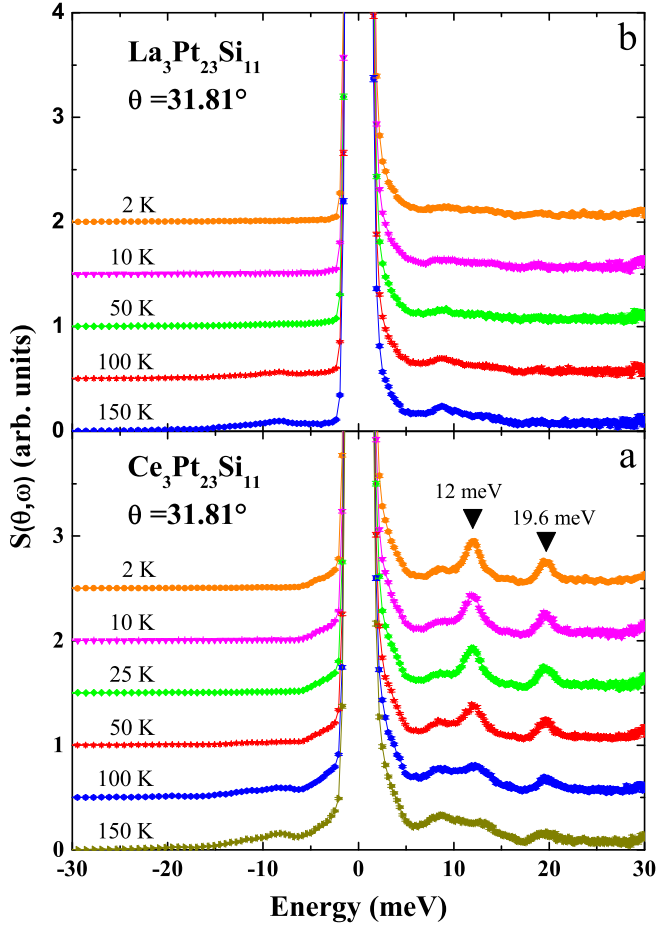}
\caption{\label{fig6} (Color online) Thermal evolution of the inelastic spectra with incident neutrons of E$_i$ = 36.7 meV (a) for Ce$_3$Pt$_{23}$Si$_{11}$, the arrows show the positions of the magnetic excitations, (b) for La$_3$Pt$_{23}$Si$_{11}$. For clarity the spectra are shifted vertically by 0.5.}
\end{figure}

\begin{figure}
\includegraphics[width=\columnwidth]{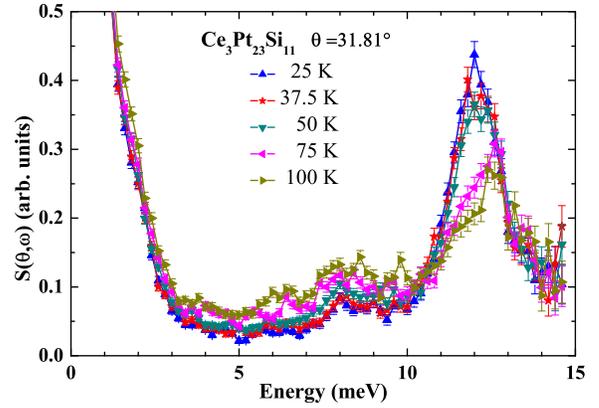}
\caption{\label{fig7} (Color online) Thermal evolution of the inelastic spectra (incident neutron energy E$_i$ = 16.6 meV) in Ce$_3$Pt$_{23}$Si$_{11}$. The non-symmetric decrease with increasing the temperature of the intensity of the peak at 12 meV reveals a phonon contribution around 12.5 - 13.5 meV.}
\end{figure}

With the purpose of comparing the magnetic scattering of the different compounds, the data obtained previously for the Ce$_3$Pt$_{23}$Si$_{11}$ and La$_3$Pt$_{23}$Si$_{11}$ samples have been treated in the same manner as those of Pr$_3$Pt$_{23}$Si$_{11}$ and Nd$_3$Pt$_{23}$Si$_{11}$. They are displayed in figures~\ref{fig6} and \ref{fig7}. At T = 2 K in the Ce$_3$Pt$_{23}$Si$_{11}$ spectrum (see figure~\ref{fig6}-a) two main inelastic structures are observed at E$_1$ = 12$\pm$0.5 meV of FWHM = 2.0 meV and at E$_2$ = 19.6$\pm$0.2 meV of FWHM = 1.7 meV respectively. Their thermal evolution is consistent with that of magnetic excitations. The excitation at 12 meV is larger than the experimental resolution. The existence of wide phonon structures around 8.8 and 13.6 meV is confirmed by the La$_3$Pt$_{23}$Si$_{11}$ spectra (see figure~\ref{fig6}-b). As in Pr- and Nd$_3$Pt$_{23}$Si$_{11}$, in the Ce compound the excitation at 12 meV is mixed with a phonon contribution. Figure~\ref{fig7} shows that the energies of the two excitations, the magnetic one at 12 meV and the phonon structure around 13.6 meV, are too close to be resolved within the experimental accuracy. Further spectra of La$_3$Pt$_{23}$Si$_{11}$ collected with an incident energy of 16.6 meV, not shown here, confirm the existence of a large phonon structure (FWHM = 2 meV) at 8.8 meV. Then the bump around 8.5 meV (FWHM = 2.4 meV) in the Ce spectra is due to phonon excitations.\\

\section{Analysis and discussion}

\begin{figure}
\includegraphics[width=\columnwidth]{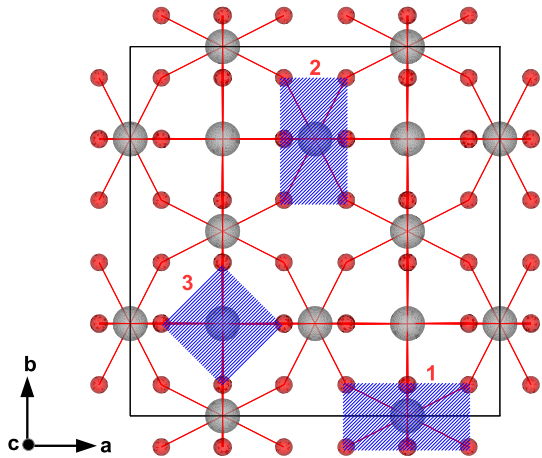}
\caption{\label{fig8} (Color online) The R[Pt8] coordinations in the phase centered cubic structure ($Fm\bar{3}m$ space group) of the R$_3$Pt$_{23}$Si$_{11}$: projection on the \textit{b,c} plane of the R[Pt8] tetragonal prisms in the unit cell. The rare earth ions are in the \textit{24d} sites and the first neighbors Pt ions in the \textit{96k} sites. The orientation of these tetragonal prisms depends on the coordinates of the six equivalent rare earth positions in the cell. For rare earths at the ($\frac{1}{4},\frac{1}{4},0$) or ($\frac{3}{4},\frac{1}{4},0$) sites the height Pt first neighbors form two rectangles lying in ($x,y$) planes at $z/a=\pm$0.0843 respectively (3). For rare earths at the sites ($\frac{1}{4},0,\frac{1}{4}$) or ($\frac{1}{4},0,\frac{3}{4}$), the Pt rectangles are lying in ($x,z$) planes at $y/a=\pm$0.0843 (1), while at the ($0,\frac{1}{4},\frac{1}{4}$) or ($0,\frac{3}{4},\frac{1}{4}$) sites they are lying in ($y,z$) planes at $x/a$=$\pm$0.0843 (2).}
\end{figure}

Despite the R$_3$Pt$_{23}$Si$_{11}$ compounds crystallize within a highly symmetric face centered cubic structure ($Fm\bar{3}m$ space group), the point symmetry of the rare earth site (24$d$) is orthorhombic (\textit{m.mm}, group $D_{2h}$). The rare earth ions are coordinated by tetragonal prisms of eight Pt atoms (in the sites 96$k$), R[Pt$8$]. As shown in figure~\ref{fig8}, the orientation of the tetragonal prism differs for the six equivalent rare earth sites within the unit cell.\\

\subsection{Orthorhombic crystal field}

Group-theory allows predicting how the degeneracy of the rare earth ion \textit{J} multiplets is lifted by the CEF interactions.~\cite{tinkham2003} At a site of orthorhombic symmetry, a complete splitting is achieved. For half integer \textit{J}, the multiplets split into $\frac{2J+1}{2}$ magnetic doublets. Therefore, within the fundamental multiplet, the maximum number of CEF transitions from the ground state is $\frac{2J+1}{2}$-1. For integer \textit{J} (non-Kramer's ions), the multiplets are split in $2J+1$ non-magnetic singlets. At most $2J$ CEF transitions can be expected from the ground state within the fundamental multiplet.\\
For Ce$^{3+}$ (\textit{J}=5/2) and Nd$^{3+}$ (\textit{J}=9/2) ions, the fundamental multiplet is decomposed into three and five magnetic doublets respectively. This is consistent with the number of magnetic excitations observed in Ce$_3$Pt$_{23}$Si$_{11}$ and Nd$_3$Pt$_{23}$Si$_{11}$. This also agrees with the value, $R\ln2$ per ion, of the magnetic entropy at the ordering temperature deduced from specific heat measurements in both compounds.\cite{opagiste2012}\\
In the case of Pr$^{3+}$ ions, the CEF interactions split the \textit{J}=4 multiplet into 9 non-magnetic singlets. This explains the low temperature Van Vleck susceptibility of Pr$_3$Pt$_{23}$Si$_{11}$. However only four transitions are observed in the neutron inelastic spectra among the eight expected. This suggests that some transitions are weak or forbidden.\\

The CEF Hamiltonian acting on a rare earth ion at a site of \textit{mmm} point group symmetry is written as:~\cite{hutchings1964}
\begin{eqnarray}
\mathcal{H}_{CEF}=&&\alpha_J(V^{0}_{2}O^{0}_{2}+V^{2}_{2}O^{2}_{2})\nonumber\\&&+\beta_J(V^{0}_{4}O^{0}_{4}+V^{2}_{4}O^{2}_{4}+V^{4}_{4}O^{4}_{4})\nonumber\\&&+\gamma_J(V^{0}_{6}O^{0}_{6}+V^{2}_{6}O^{2}_{6}+V^{4}_{6}O^{4}_{6}+V^{6}_{6}O^{6}_{6})
\label{eq:one}
\end{eqnarray}
The $O^{m}_{l}$ are the Stevens operators.\cite{stevens1952} $\alpha_J$, $\beta_J$ and $\gamma_J$ are the 2nd-, 4th- and 6th-order Stevens coefficients, the value of which depends on the rare earth. The $V^{m}_{l}$ are the CEF parameters that depend on the surroundings. The number of CEF parameters is too large, nine parameters for Pr and Nd and five for Ce ( $\gamma_J$ = 0), to be determined from the neutron spectroscopy only. Many sets of CEF parameter can be found that comply with the energy scheme deduced from the NS spectra. However, CEF-based calculations show that only very few are simultaneously consistent with the magnetic properties. In particular, calculations should account for the thermal variation of the first-order magnetic susceptibility, for the anisotropy, the field and temperature dependence of the magnetization.

In an orthorhombic system, the second-order CEF parameters can be deduced from the anisotropy of the magnetic susceptibility measured on single crystals along the three axes of the orthorhombic structure.~\cite{fillion1984} In the present case, these parameters cannot be obtained since the symmetry of the crystal is cubic and the first-order magnetic susceptibility isotropic. Therefore an alternative approach has to be implemented in order to determine the whole sets of CEF parameters.\\

\subsection{Genetic algorithm: search for CEF parameters}

As no single crystal measurements on Pr$_3$Pt$_{23}$Si$_{11}$ and Nd$_3$Pt$_{23}$Si$_{11}$ are available, the determination of the nine CEF parameters for these compounds appears hopeless. 
To this regard, the case of Ce$_3$Pt$_{23}$Si$_{11}$ seems more favorable. Indeed only five CEF parameters are to be determined and more experimental data are available: the spin arrangement in the ordered phase, the value of the moment at 100 mK, magnetic measurements on single crystal in both the paramagnetic and ferromagnetic phases.\\

In the search for the CEF parameters of Ce$_3$Pt$_{23}$Si$_{11}$, we proceed first by effectively exploring the space formed by the sets of $V^{m}_{l}$ parameters that yield an energy level schema compatible with the energies of the experimental excitations, E$_1$ = 139 K and E$_2$ = 227.5 K. In this optimization problem, many sets of values can be found because of the small number of imposed constraints. We therefore developed a numerical program based on genetic algorithms (GA), which constitutes the most adapted approach for solving a non unique solution problem.~\cite{goldberg1989} In our implementation the GA population of candidate solutions was limited to 100 and its evolution was studied during 400 generations.
The range of the $V_{lm}$ parameters were bounded to a reasonable energy range, [-500 K, 500 K], and coded on 32 bit words, each word representing a chromosome and each bit a gene in the GA terminology. The quality of a candidate solution was evaluated through its fitness function, which, in our case, measures the sum of the squared differences between the calculated energy gaps and the expected ones. At each generation, the GA population was submitted to crossover operations between chromosomes with a rate of 90\%, allowing to explore locally the parameter space. Mutation operations were also applied with a rate of 0.2\%, allowing to make huge leaps and to capture other possible solutions. Only the best candidate solution was kept. We then proceeded to a local minimization based on a simplex method to refine the $V_{lm}$ values, thus increasing their precision.~\cite{nelder1965}\\

This procedure was iteratively used to sample the subspace of CEF parameters that comply with the energy level schema. The compliance of these sets with the magnetic properties has yet to be considered. An important criteria is the value of the ordered magnetic moment at low temperature. An experimental value of 1.2$\pm0.2 \mu_B$ was deduced from neutron diffraction measurements at 100 mK.~\cite{opagiste2011} Among the solutions of the genetic algorithm, only the sets that give a value for the moment of the fundamental doublet below 1.7 $\mu_B$/Ce were retained for the next step.\\

\subsection{Molecular field model}

In order to compare the magnetic experimental data with CEF based calculations, a model adapted to the particular crystalline structure is required. The unit cell of the R$_3$Pt$_{23}$Si$_{11}$ compounds contains six equivalent rare-earth sites. The simplest microscopic model has to account for the effects of the CEF, the molecular exchange field $\bm{H}_m = n \bm{M}$ ($n$ is the molecular field constant and $\bm{M}$ the magnetization) and the applied magnetic field $\bm{H}$ (both $\bm{H}_m$ and $\bm{H}$ in Tesla unit). The Hamiltonian describing site $i$ then reads as :
\begin{equation}
\label{hamilt}
\mathcal{H}_{i} = {\mathcal{H}_{CEF}}_i + \mu_B\,{g_J} \bm{H} \cdot \bm{J}_i + \mu_B\,{g_J} {\bm{H}_m}_i \cdot \bm{J}_i
\end{equation}

As the system orders ferromagnetically, the molecular and applied magnetic fields can be merged into a total field $\bm{H}_T$, identical on all sites, in both the paramagnetic and ordered states. 
\begin{equation}
\mathcal{H}_{i} = {\mathcal{H}_{CEF}}_i + \mu_B\,{g_J} \bm{H}_T \cdot \bm{J}_i
\end{equation}

The orientation of the orthorhombic axes varying from site to site (see figure 8), the six ions cannot be described by the same CEF Hamiltonian $\mathcal{H}_{CEF}$. In order to use the same CEF expression for all sites, one would need six, symmetry equivalent, sets of CEF parameters. An alternative way consists in using the same CEF Hamiltonian, with a single set of CEF parameters describing a reference site (with index 0). Any other site $i$ can be brought in coincidence with this reference by application of a transformation $T_i$ consisting in rotations from the cubic point group of the crystallographic cell. Then, instead of treating the Hamiltonian $\mathcal{H}_{i}$ in order to obtain the statistical moment $\bm{m}_i$, one can solve the question using the Hamiltonian $\mathcal{H}_{0}$ where the total field is replaced by $T(\bm{H}_T)$, which yields a magnetic moment $\bm{m}$. The magnetic moment $\bm{m}_i$ is then obtained by rotating $\bm{m}$ back to the original orientation of site $i$ : $\bm{m}_i = T^{-1}(\bm{m})$. In this way, all the six magnetic moments in the unit cell can be computed, which allows to define the magnetization $\bm{M}$ and, therefore, the molecular field $\bm{H}_m = n \bm{M}$ for the next iteration... until convergence.\\

We can now proceed, for a given set of CEF parameters, to the calculations of the thermal variation of the susceptibility. These CEF parameters are then optimized by minimizing a $\chi^2$ that accounts for the deviations from both the experimental low temperature susceptibility and energy scheme. The optimized set is qualified only if the associated deviations are within the experimental errors.\\
In a last step, we checked whether the low temperature transition probabilities between the CEF ground state and the two excited doublets are compatible with the intensity of the observed magnetic transitions (see figure~\ref{fig6}). Gaussian fits to the 2 K spectra show that the intensity ratio, I(E$_2$)/I(E$_1$), is of the order of 0.4. Among the qualified sets of CEF parameters, only one nicely fulfills this last criteria.
\begin{table}
\caption{\label{table1}{Ce$_3$Pt$_{23}$Si$_{11}$: Eigenstates and eigenvalues of the CEF Hamiltonian, $\mathcal{H}_{0}$ for the reference site, calculated with the set of CEF parameters $V^{0}_{2}$ = 182 K, $V^{2}_{2}$ = 133.9 K, $V^{0}_{4}$ = -1.7 K, $V^{2}_{4}$ = -17.2 K and $V^{4}_{4}$ = 285.2 K.}}
\begin{ruledtabular}
\begin{tabular}{cc}
Eigenstates & Eigenvalues (K)\\
\colrule
$+0.950\left|\pm\frac{5}{2}\right\rangle-0.305\left|\mp\frac{3}{2}\right\rangle+0.064\left|\pm\frac{1}{2}\right\rangle$ & -121.8\\
$-0.207\left|\pm\frac{5}{2}\right\rangle-0.462\left|\mp\frac{3}{2}\right\rangle+0.862\left|\pm\frac{1}{2}\right\rangle$ & +16.8\\
$\mp0.233\left|\pm\frac{5}{2}\right\rangle\mp0.833\left|\mp\frac{3}{2}\right\rangle\mp0.502\left|\pm\frac{1}{2}\right\rangle$ & +105\\
\end{tabular}
\end{ruledtabular}
\end{table}

\begin{figure}
\includegraphics[width=\columnwidth]{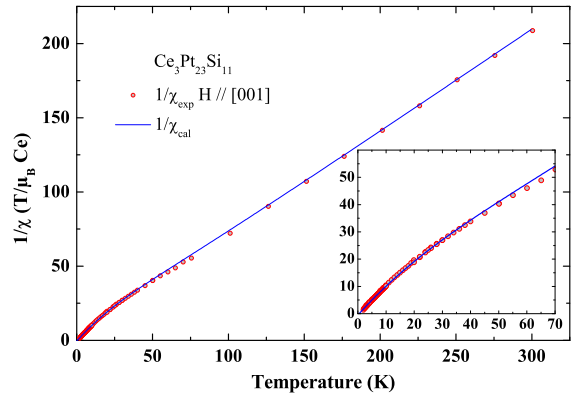}
\caption{\label{fig9} (Color online) Calculated (lines) and experimental (dots) thermal variation of the inverse susceptibility. The inset details the low temperature curves. The calculations are performed with the set of CEF parameters given in Table~\ref{table1} and the molecular field constant \textit{n} = 0.5 T/$\mu_B$. The systematic error in the experimental susceptibility curve is of the order of 2\%.}
\end{figure}

This set of CEF parameters is: $V^{0}_{2}$ = 182 K, $V^{2}_{2}$ = 133.9 K, $V^{0}_{4}$ = -1.7 K, $V^{2}_{4}$ = -17.2 K and $V^{4}_{4}$ = 285.2 K. The eigenstates and eigenvalues of the CEF Hamiltonian are given in table~\ref{table1}. An exchange constant of \textit{n} = 0.5 T/$\mu_B$ has been determined from the value of the inverse CEF susceptibility at \textit{T}$_C$ = 0.44 K. The calculated inverse susceptibility is compared in figure~\ref{fig9} with the experimental data.\\

\begin{figure}
\includegraphics[width=\columnwidth]{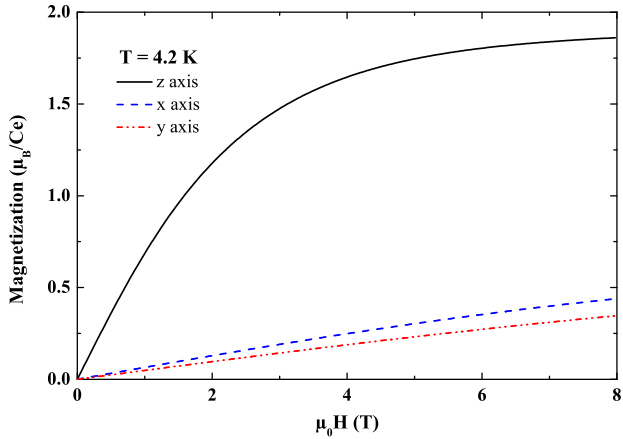}
\caption{\label{fig10} (Color online) Calculated magnetization curves for fields applied along the x, y and z axes and at T = 4.2 K, for a Ce$^{3+}$ at the reference orthorhombic 0 site, under the influence of the CEF defined by the parameters in Table~\ref{table1}.}
\end{figure}

\subsection{Analysis of the magnetization processes}

\begin{figure}
\includegraphics[width=\columnwidth]{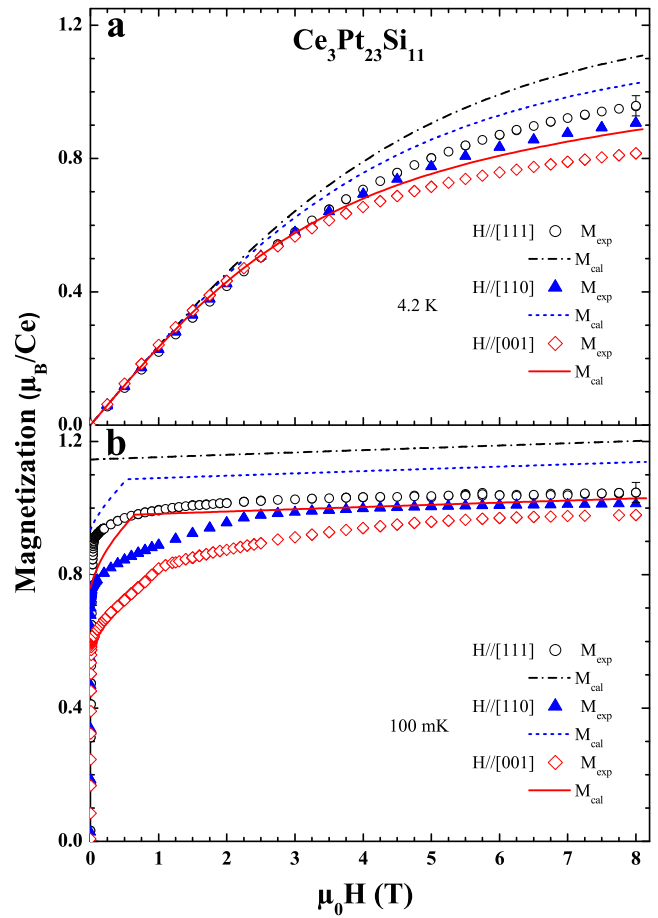}
\caption{\label{fig11} (Color online) Calculated (lines) and experimental (dots) magnetization curves for fields applied along the threefold, twofold and fourfold axes of the cubic cell at (a) 4.2 K (b) 100 mK. The relative experimental error on the magnetization value is estimated of the order of 3\%. The calculations are performed with the set of CEF parameters given in Table ~\ref{table1} and a molecular field constant \textit{n} = 0.5 T/$\mu_B$.}
\end{figure}

As shown in figure~\ref{fig9} the calculated and experimental thermal variations of the inverse susceptibility are in good agreement. The isotropy of the first-order magnetic susceptibility conceals the large magnetic anisotropy, resulting from the orthorhombic CEF at the Ce$^{3+}$ site.
This anisotropy is illustrated in Figure~\ref{fig10}, which shows the magnetization curves calculated, at 4.2 K for the reference Ce$^{3+}$ site, for fields applied along the \textit{x}, \textit{y} and \textit{z} axes. It appears that a significant magnetization is obtained only along the \textit{z} axis. In the specific crystal structure of the R$_3$Pt$_{23}$Si$_{11}$ compounds, at each R site this axis is in coincidence with one of the fourfold axes of the cubic cell. In case of a magnetic field applied along a fourfold axis of the cubic cell, only one third of the Ce$^{3+}$ ions substantially contribute to the magnetization. Reciprocally, for a field applied along a threefold axis of the cubic cell, all Ce$^{3+}$ ions contribute to the magnetization via a $1/\sqrt{3}$ projection ratio.\\
Indeed, the magnetization measurements (see figure~\ref{fig11}) show that the effective easy axis is the threefold one, in agreement with the CEF-based calculations for the cubic system. Figure~\ref{fig11}-a compares the calculated magnetization at 4.2 K with the experimental curves measured with fields applied along the three main symmetry axes of the cube. The calculations reproduce qualitatively well the experimental magnetization curves and, as expected, they find the easy and hard magnetization axes, along the threefold and fourfold axes respectively. However, above 2 T, the calculated magnetization is systematically larger than the experimental one by 8 to 13\%. In the ferromagnetic phase (see figure~\ref{fig11}-b), the calculations confirm an easy magnetization axis along the threefold direction, in agreement with the experiments. At 100 mK, the experimental curves along the [110] and [001] axes show kinks at about 2.5 and 1 T respectively. Calculations qualitatively reproduce these discontinuities, which correspond to the end of the rotation of the moments towards the direction of the applied field. The magnetization is again calculated larger than the experimental values.

The used model handles only the CEF effects on the magnetic properties of the Ce ions. Actually, there exist other contributions to the magnetization, from the matrix and/or from the conduction electrons. They may become sizable with respect to that of the rare earth ions in this rather diluted system. Depending on their signs they can lead to an effective reduction of the magnetic signal. In most cases, these contributions can be well accounted for by the signal of the isostructural La compound. In the present case the value of the diamagnetic susceptibility in La$_3$Pt$_{23}$Si$_{11}$, $\chi\approx-2\times10^{-5}$ $\mu_B/T$, cannot explain alone the disagreement between calculations and experimental data. An other contribution, due the local polarization of the conduction electrons by the rare earth ions, can lead to a reduction of the moment. As Ce$_3$Pt$_{23}$Si$_{11}$ shows no evidence of Kondo coupling, this local polarization is opposite to that of the rare earth. According to the calculated moment for the Ce ions at 100 mK, $\muµ_{cal}$ = 1.57 $\muµ_B$, and the value of the moment determined by neutron diffraction, $\muµ$ = 1.2$\pm0.2~\muµ_B$, this polarization would be of the order of 0.2-0.4 $\mu_B$ per Ce site. The existence of such a polarization could be confirmed experimentally by polarized neutron diffraction. It should be emphasized, however, that in rare earth compounds, higher-orders effects, such as magnetostriction, may manifest under large magnetic fields. Those, which are not accounted for in the Hamiltonian of Eq.~\ref{hamilt}, may also explain the excess in the calculated magnetization.\\

\section{Conclusion}

The number of magnetic excitations in the neutron spectroscopy spectra of the Ce, Pr and Nd compounds in the R$_3$Pt$_{23}$Si$_{11}$ series, is consistent with the orthorhombic symmetry at the rare earth site. Unfortunately, for Pr- and Nd$_3$Pt$_{23}$Si$_{11}$ the number of CEF parameters in the Hamiltonian is too large to allow for their unambiguous determination. This impedes further analysis of the magnetic properties. In the case of Ce$_3$Pt$_{23}$Si$_{11}$, the reduced number of CEF parameters (5) and the availability of comprehensive experimental data alleviate the difficulty. To determine the CEF at the cerium site, a protocol has been developed. First, a collection of sets of CEF parameters, that comply with the neutron energy level schema, is obtained by a genetic algorithm method. Second, using a microscopic model adapted to the special crystal structure of the R$_3$Pt$_{23}$Si$_{11}$, the initial sets are self-consistently optimized in order to simultaneously describe the CEF energy schema and the first-order magnetic susceptibility. Last, the transition probabilities between the ground and the excited CEF states, as derived from the intensities in the NS spectra, are confronted with those computed for the optimized sets of CEF parameters. This allows to identify an unique set of optimized parameters. Calculations show that this CEF is responsible for a large magnetic anisotropy of the Ce ions. Combining the response of the 6 Ce sites in the cubic crystallographic shell, the magnetization processes in the paramagnetic range are well reproduced. The large orthorhombic anisotropy is directly responsible for the easy threefold and hard fourfold axes of this cubic system. Beyond the Ce case, this should stand for all elements in the R$_3$Pt$_{23}$Si$_{11}$ series. The ferromagnetic phase of Ce$_3$Pt$_{23}$Si$_{11}$ is governed by the competition between the local orthorhombic anisotropy and the exchange couplings, resulting in a magnetization along a threefold axis. Calculations including an applied magnetic field allow to interpret the evolution of the magnetization at very low temperature : the compromise between the anisotropy and the total field results in progressive rotations of the magnetic moments. Ce$_3$Pt$_{23}$Si$_{11}$ is thus an interesting model compound that helps understanding the behavior of systems where conflicting local anisotropies are forced to cooperate within the overall higher symmetry of the crystal.\\

\begin{acknowledgments}
Authors greatly acknowledge R. Haettel from the Institut N\'eel and O. Meulien from the Institut Laue-Langevin for their technical assistance.
\end{acknowledgments}

\bibliography{Neutron_R3Pt23Si11compounds}

\begin{thebibliography}{12}%
\makeatletter
\providecommand \@ifxundefined [1]{%
 \@ifx{#1\undefined}
}%
\providecommand \@ifnum [1]{%
 \ifnum #1\expandafter \@firstoftwo
 \else \expandafter \@secondoftwo
 \fi
}%
\providecommand \@ifx [1]{%
 \ifx #1\expandafter \@firstoftwo
 \else \expandafter \@secondoftwo
 \fi
}%
\providecommand \natexlab [1]{#1}%
\providecommand \enquote  [1]{``#1''}%
\providecommand \bibnamefont  [1]{#1}%
\providecommand \bibfnamefont [1]{#1}%
\providecommand \citenamefont [1]{#1}%
\providecommand \href@noop [0]{\@secondoftwo}%
\providecommand \href [0]{\begingroup \@sanitize@url \@href}%
\providecommand \@href[1]{\@@startlink{#1}\@@href}%
\providecommand \@@href[1]{\endgroup#1\@@endlink}%
\providecommand \@sanitize@url [0]{\catcode `\\12\catcode `\$12\catcode
  `\&12\catcode `\#12\catcode `\^12\catcode `\_12\catcode `\%12\relax}%
\providecommand \@@startlink[1]{}%
\providecommand \@@endlink[0]{}%
\providecommand \url  [0]{\begingroup\@sanitize@url \@url }%
\providecommand \@url [1]{\endgroup\@href {#1}{\urlprefix }}%
\providecommand \urlprefix  [0]{URL }%
\providecommand \Eprint [0]{\href }%
\@ifxundefined \urlstyle {%
  \providecommand \doi  [0]{\begingroup \@sanitize@url \@doi}%
  \providecommand \@doi [1]{\endgroup \@@startlink {\doibase
  #1}doi:\discretionary {}{}{}#1\@@endlink }%
}{%
  \providecommand \doi  [0]{doi:\discretionary{}{}{}\begingroup
  \urlstyle{rm}\Url }%
}%
\providecommand \doibase [0]{http://dx.doi.org/}%
\providecommand \Doi [0]{\begingroup \@sanitize@url \@Doi }%
\providecommand \@Doi  [1]{\endgroup\@@startlink{\doibase#1}\@@Doi}%
\providecommand \@@Doi [1]{#1\@@endlink}%
\providecommand \selectlanguage [0]{\@gobble}%
\providecommand \bibinfo  [0]{\@secondoftwo}%
\providecommand \bibfield  [0]{\@secondoftwo}%
\providecommand \translation [1]{[#1]}%
\providecommand \BibitemOpen [0]{}%
\providecommand \bibitemStop [0]{}%
\providecommand \bibitemNoStop [0]{.\EOS\space}%
\providecommand \EOS [0]{\spacefactor3000\relax}%
\providecommand \BibitemShut  [1]{\csname bibitem#1\endcsname}%
\bibitem [{\citenamefont {Opagiste}\ and\ \citenamefont
  {Gal\'era}(2012)}]{opagiste2012}%
  \BibitemOpen
  \bibfield  {author} {\bibinfo {author} {\bibfnamefont {C.}~\bibnamefont
  {Opagiste}}\ and\ \bibinfo {author} {\bibfnamefont {R.-M.}\ \bibnamefont
  {Gal\'era}},\ }\href@noop {} {\bibfield  {journal} {\bibinfo  {journal}
  {Journal of Alloys and Compounds},\ }\textbf {\bibinfo {volume} {541}},\
  \bibinfo {pages} {403 } (\bibinfo {year} {2012})},\ ISSN \bibinfo {issn}
  {0925-8388}\BibitemShut {NoStop}%
\bibitem [{\citenamefont {Opagiste}\ and\ \citenamefont
  {Gal\'era}(2014)}]{opagiste2014}%
  \BibitemOpen
  \bibfield  {author} {\bibinfo {author} {\bibfnamefont {C.}~\bibnamefont
  {Opagiste}}\ and\ \bibinfo {author} {\bibfnamefont {R.-M.}\ \bibnamefont
  {Gal\'era}},\ }\href@noop {} {\bibfield  {journal} {\bibinfo  {journal}
  {Journal of Magnetism and Magnetic Materials},\ }\textbf {\bibinfo {volume}
  {357}},\ \bibinfo {pages} {13 } (\bibinfo {year} {2014})},\ ISSN \bibinfo
  {issn} {0304-8853}\BibitemShut {NoStop}%
\bibitem [{\citenamefont {Opagiste}\ \emph {et~al.}(2015)\citenamefont
  {Opagiste}, \citenamefont {Barbier}, \citenamefont {Haettel},\ and\
  \citenamefont {Gal\'era}}]{opagiste2015}%
  \BibitemOpen
  \bibfield  {author} {\bibinfo {author} {\bibfnamefont {C.}~\bibnamefont
  {Opagiste}}, \bibinfo {author} {\bibfnamefont {C.}~\bibnamefont {Barbier}},
  \bibinfo {author} {\bibfnamefont {R.}~\bibnamefont {Haettel}}, \ and\
  \bibinfo {author} {\bibfnamefont {R.-M.}\ \bibnamefont {Gal\'era}},\ }\Doi
  {http://dx.doi.org/10.1016/j.jmmm.2014.11.070} {\bibfield  {journal}
  {\bibinfo  {journal} {Journal of Magnetism and Magnetic Materials},\ }\textbf
  {\bibinfo {volume} {378}},\ \bibinfo {pages} {402 } (\bibinfo {year}
  {2015})},\ ISSN \bibinfo {issn} {0304-8853}\BibitemShut {NoStop}%
\bibitem [{\citenamefont {Tursina}\ \emph {et~al.}(2002)\citenamefont
  {Tursina}, \citenamefont {Gribanov}, \citenamefont {Seropegin}, \citenamefont
  {Kuyukov},\ and\ \citenamefont {Bodak}}]{tursina2002}%
  \BibitemOpen
  \bibfield  {author} {\bibinfo {author} {\bibfnamefont {A.~I.}\ \bibnamefont
  {Tursina}}, \bibinfo {author} {\bibfnamefont {A.~V.}\ \bibnamefont
  {Gribanov}}, \bibinfo {author} {\bibfnamefont {Y.~D.}\ \bibnamefont
  {Seropegin}}, \bibinfo {author} {\bibfnamefont {K.~V.}\ \bibnamefont
  {Kuyukov}}, \ and\ \bibinfo {author} {\bibfnamefont {O.~I.}\ \bibnamefont
  {Bodak}},\ }\Doi {DOI: 10.1016/S0925-8388(02)00758-2} {\bibfield  {journal}
  {\bibinfo  {journal} {Journal of Alloys and Compounds},\ }\textbf {\bibinfo
  {volume} {347}},\ \bibinfo {pages} {121 } (\bibinfo {year}
  {2002})}\BibitemShut {NoStop}%
\bibitem [{\citenamefont {Opagiste}\ \emph {et~al.}(2013)\citenamefont
  {Opagiste}, \citenamefont {Jackson}, \citenamefont {Gal\'era}, \citenamefont
  {Lhotel}, \citenamefont {Paulsen},\ and\ \citenamefont
  {Ouladdiaf}}]{opagiste2013}%
  \BibitemOpen
  \bibfield  {author} {\bibinfo {author} {\bibfnamefont {C.}~\bibnamefont
  {Opagiste}}, \bibinfo {author} {\bibfnamefont {M.}~\bibnamefont {Jackson}},
  \bibinfo {author} {\bibfnamefont {R.-M.}\ \bibnamefont {Gal\'era}}, \bibinfo
  {author} {\bibfnamefont {E.}~\bibnamefont {Lhotel}}, \bibinfo {author}
  {\bibfnamefont {C.}~\bibnamefont {Paulsen}}, \ and\ \bibinfo {author}
  {\bibfnamefont {B.}~\bibnamefont {Ouladdiaf}},\ }\href@noop {} {\bibfield
  {journal} {\bibinfo  {journal} {Journal of Magnetism and Magnetic
  Materials},\ }\textbf {\bibinfo {volume} {340}},\ \bibinfo {pages} {46 }
  (\bibinfo {year} {2013})},\ ISSN \bibinfo {issn} {0304-8853}\BibitemShut
  {NoStop}%
\bibitem [{\citenamefont {Opagiste}\ \emph {et~al.}(2011)\citenamefont
  {Opagiste}, \citenamefont {Gal\'era}, \citenamefont {Amara}, \citenamefont
  {Paulsen}, \citenamefont {Rols},\ and\ \citenamefont
  {Ouladdiaf}}]{opagiste2011}%
  \BibitemOpen
  \bibfield  {author} {\bibinfo {author} {\bibfnamefont {C.}~\bibnamefont
  {Opagiste}}, \bibinfo {author} {\bibfnamefont {R.-M.}\ \bibnamefont
  {Gal\'era}}, \bibinfo {author} {\bibfnamefont {M.}~\bibnamefont {Amara}},
  \bibinfo {author} {\bibfnamefont {C.}~\bibnamefont {Paulsen}}, \bibinfo
  {author} {\bibfnamefont {S.}~\bibnamefont {Rols}}, \ and\ \bibinfo {author}
  {\bibfnamefont {B.}~\bibnamefont {Ouladdiaf}},\ }\href@noop {} {\bibfield
  {journal} {\bibinfo  {journal} {Physical Review B},\ }\textbf {\bibinfo
  {volume} {84}},\ \bibinfo {pages} {134401} (\bibinfo {year}
  {2011})}\BibitemShut {NoStop}%
\bibitem [{\citenamefont {Tinkham}(2003)}]{tinkham2003}%
  \BibitemOpen
  \bibfield  {author} {\bibinfo {author} {\bibfnamefont {M.}~\bibnamefont
  {Tinkham}},\ }\href@noop {} {\emph {\bibinfo {title} {Group Theory and
  Quantum Mechanics}}}\ (\bibinfo  {publisher} {Dover Publications, N.Y.},\
  \bibinfo {year} {2003})\BibitemShut {NoStop}%
\bibitem [{\citenamefont {Hutchings}(1964)}]{hutchings1964}%
  \BibitemOpen
  \bibfield  {author} {\bibinfo {author} {\bibfnamefont {M.~T.}\ \bibnamefont
  {Hutchings}},\ }\href@noop {} {\bibfield  {journal} {\bibinfo  {journal}
  {Solid State Physics},\ }\textbf {\bibinfo {volume} {16}},\ \bibinfo {pages}
  {227} (\bibinfo {year} {1964})}\BibitemShut {NoStop}%
\bibitem [{\citenamefont {Stevens}(1952)}]{stevens1952}%
  \BibitemOpen
  \bibfield  {author} {\bibinfo {author} {\bibfnamefont {K.~W.~H.}\
  \bibnamefont {Stevens}},\ }\href@noop {} {\bibfield  {journal} {\bibinfo
  {journal} {Proc. Phys. Soc. A},\ }\textbf {\bibinfo {volume} {65}},\ \bibinfo
  {pages} {209} (\bibinfo {year} {1952})}\BibitemShut {NoStop}%
\bibitem [{\citenamefont {Fillion}\ \emph {et~al.}(1984)\citenamefont
  {Fillion}, \citenamefont {Gignoux}, \citenamefont {Givord},\ and\
  \citenamefont {Lemaire}}]{fillion1984}%
  \BibitemOpen
  \bibfield  {author} {\bibinfo {author} {\bibfnamefont {G.}~\bibnamefont
  {Fillion}}, \bibinfo {author} {\bibfnamefont {D.}~\bibnamefont {Gignoux}},
  \bibinfo {author} {\bibfnamefont {F.}~\bibnamefont {Givord}}, \ and\ \bibinfo
  {author} {\bibfnamefont {R.}~\bibnamefont {Lemaire}},\ }\Doi
  {http://dx.doi.org/10.1016/0304-8853(84)90060-X} {\bibfield  {journal}
  {\bibinfo  {journal} {Journal of Magnetism and Magnetic Materials},\ }\textbf
  {\bibinfo {volume} {44}},\ \bibinfo {pages} {173 } (\bibinfo {year}
  {1984})},\ ISSN \bibinfo {issn} {0304-8853}\BibitemShut {NoStop}%
\bibitem [{\citenamefont {Goldberg}(1989)}]{goldberg1989}%
  \BibitemOpen
  \bibfield  {author} {\bibinfo {author} {\bibfnamefont {D.}~\bibnamefont
  {Goldberg}},\ }\href@noop {} {\emph {\bibinfo {title} {Genetic Algorithms in
  Search, Optimization, and Machine Learning}}}\ (\bibinfo  {publisher}
  {Addison-Wesley Professional},\ \bibinfo {year} {1989})\BibitemShut {NoStop}%
\bibitem [{\citenamefont {Nelder}\ and\ \citenamefont
  {Mead}(1965)}]{nelder1965}%
  \BibitemOpen
  \bibfield  {author} {\bibinfo {author} {\bibfnamefont {J.}~\bibnamefont
  {Nelder}}\ and\ \bibinfo {author} {\bibfnamefont {R.}~\bibnamefont {Mead}},\
  }\href@noop {} {\bibfield  {journal} {\bibinfo  {journal} {Computer
  Journal},\ }\textbf {\bibinfo {volume} {7}},\ \bibinfo {pages} {308}
  (\bibinfo {year} {1965})}\BibitemShut {NoStop}%
\end{thebibliography}%

\end{document}